\def\Tr{{\rm Tr}}
\def\hxp{\hat X_+}
\def\hxm{\hat X_-}
\def\rhc{{\check{\rho}}}
\def\fing{{\bf g}}
\def\gh{{\bf \hat g}}
\def\beq{\begin{equation}}
\def\eeq{\end{equation}}
\newtheorem{lemma}{Lemma}
\def\beqa{\begin{eqnarray}}
\def\eeqa{\end{eqnarray}}

\def\ad{{\rm ad}}
\def\faz{\hat F(\alpha,z)}

\documentstyle[11pt,amssymbols,a4]{article}

\let\ssection=\section
\renewcommand{\section}{\setcounter{equation}{0}\ssection}

\begin{document}
\begin{titlepage}
\hfill Imperial/TP/92-93/30

\hfill {\tt hep-th 9304156}

\vskip 3cm
\begin{center}
\huge
Aspects of Non-Abelian Toda Theories
\vskip 2cm
\Large
Jonathan Underwood
\vskip 2cm\large
{\it Blackett Laboratory, Imperial College, London SW7 2AZ, United Kingdom.}
\normalsize
\vskip 2cm
\end{center}
\begin{abstract}
We present a definition of the non-abelian generalisations of affine
Toda theory related from the outset to vertex operator constructions
of the corresponding Kac-Moody algebra $\gh$. Reuslts concerning
conjugacy classes of the Weyl group of the finite Lie algebra
$\fing$ to embeddings of $A_1$ in $\fing$ are used both to present
the theories, and to elucidate their
soliton spectrum. We confirm the conjecture of
\cite{OSU93} for the soliton specialisation of the Leznov-Saveliev
solution. The energy-momentum tensor of such theories
is shown to split into a total derivative part and a part dependent
only on the free fields which appear in the general solution, and
vanish for the soliton solutions. Analogues are provided of the
results known for the classical solitons of abelian Toda theories.
\end{abstract}

\end{titlepage}

\bibliographystyle{unsrt}
\section{Introduction}

The abelian Toda field theories associated with an affine Kac-Moody
algebra are now well established in the literature as examples of
integrable models. They exhibit a variety of interesting structures,
and their zero-curvature formulation (see e.g. \cite{LS92}) allows
the use of powerful techniques from the theory of Lie algebras.

The theories can also be presented from a Lagrangian point of view,
the Lagrangian being the usual bosonic kinetic term minus a potential,
exponential in the field. This is the natural form if we want to
calculate the canonical energy-momentum tensor, which has been shown
\cite{OTU93}
to have some interesting properties. We can also perform the
standard perturbative canonical quantisation of the theory by treating terms
higher than quadratic in the potential as an interaction between the
so-called Toda particles. These are the quanta of the modes of the `free'
Lagrangian which remains. The masses of these
particle have an elegant algebraic origin (see e.g. \cite{FLO91}).
Their couplings also permit a Lie algebraic description, given by
Dorey's fusing rule \cite{Do91}, \cite{FLO91}. Because the theory is
integrable and two-dimensional the S-matrix for the particles has to
satisfy a very restrictive set of constraints. A solution to this
problem has been conjectured and discussed by a number of authors
\cite{Do92}, \cite{DGZ92}, \cite{BCDS90}, \cite{FO92}.

Using the grading structure of the underlying Lie algebra the
equations of motion can be completely integrated to find a
general solution \cite{LS83}, \cite{LS92} for the matrix elements of
the dynamical variable in some particular representations. This solution is
given
in terms of a free field, belonging to the same group or algebra as
the dynamical field. Using this solution it can be shown
\cite{OTU93} that the energy-momentum tensor is a sum of two parts;
a total derivative, which can be integrated to yield a surface term,
and a piece dependent only on the free fields.

Because the theories possess an infinite set of degenerate vacua we
expect to find soliton solutions interpolating them. This
possibility was first discussed by Hollowood \cite{Ho92} for the
theories based on $\hat A_r$. He used the Hirota method, a technique
later used by \cite{MM92}, \cite{ACFGZ92} to develop the
solutions for all of the theories. This technique has the
disadvantage that the solutions are not given in terms of a general
formula, rather they are iteratively defined. This tends to make
calculation of the mass spectrum rather tricky, as well as the
evaluation of the so-called topological charge which is the change
in the solution between $x=-\infty$ and $x=\infty$.

A parallel approach was developed in \cite{OTU93}, where a
specialisation of the general solution which leads to the N-soliton
solution had been given. This allowed discussion of the spectra of
conserved charges without having to find the explicit
solution for a particular theory. It also revealed a surprising
connection with the principal vertex operator construction of an
affine algebra. (A restricted solution of the $\hat A_1$ theory which contains
the soliton solutions was discussed by \cite{BB92}. These authors
used an  intuitively attractive `dressing transformation' method and
also noted  a
connection with vertex operators.)
The mass formula obtained by the methods of \cite{OTU93} reveals that the
coupling constant must
be imaginary for them to be positive. With this understanding the
masses are found to be proportional to the masses of the particles
in the Toda theory associated with the dual Lie algebra. This bears
a resemblance to duality conjectures in the theory of  self-dual Yang-Mills
monopoles, where technical difficulties appear to have prevented
their resolution.

The present work arises from a conjecture in \cite{OSU93} for the
solitonic specialisation of the general solution of the generalised
or non-abelian Toda theories defined in \cite{LS83}. Little progress
has been made in the understanding of these theories, principally on
account of the complexity of their original presentation. Here we
show that there is a description which allows us to make further
progress with them.

The structure of this paper is as follows. In section 2 we review
the Leznov-Saveliev presentation and solution of a Toda-type system,
which depends on the definition of some particular elements of a
graded Lie algebra. In section 3 we describe the original definition
of the non-abelian affine Toda theories based on the work of
\cite{LS83}. By analogy with the abelian Toda case \cite{OTU93} we
give an alternative characterisation of these theories, and discuss
the means by which it can be related to the original one. Section 4
is devoted to discussion of the abelianisation procedure for
deducing the conserved charges of these theories, and we produce an
explicit formula for the lowest one. The purpose of section 5 is to
demonstrate the splitting of the energy-momentum tensor, one part
being equal to the conserved charge calculated in the previous
section, and the other a topological `improvement' term. Section 6
uses this expression to calculate the energy and momentum of the
N-soliton solution. Finally we indicate directions for future
progress in section 7.
\section{Leznov-Saveliev Solution}

In this section we present the most general definition of a Toda
theory as a zero-curvature condition, and its solution due to Leznov
and Saveliev \cite{LS83}. Our treatment and notation will be very
much along the lines of this reference. For a detailed discussion of
many applications of group-theory to integrable systems we refer to
the monograph \cite{LS92}

\subsection{Zero-Curvature Presentation of Toda Systems}
Let us establish some notation. To define a Toda theory we need a
Lie group, $\Gamma$, whose Lie algebra $\gamma$ (assumed simple,
semisimple algebras are a trivial generalisation) has a gradation over
the integers. We label the graded subspaces
\beq\gamma=\bigoplus_{p\in\Bbb Z}\gamma_p.\label{decomp}\eeq It is
occasionally helpful to require that $I(\gamma_p)=\gamma_{-p}$,
where $I$ is the Chevalley involution with respect to a Cartan
subalgebra contained in $\gamma_0$. The Toda
field, $h=u^2$, lies in $\Gamma_0$, the exponential of the
subalgebra $\gamma_0$. Define also two elements $X_\pm\in\gamma_{\pm
1}$. Working in light-cone coordinates $x^\pm=(t\pm x)/\sqrt 2$, the
connection with components \beqa A_+ & = &
u^{-1}\partial_+u+uX_+u^{-1}\nonumber\\ A_- & =& -\partial_-uu^{-1}
+u^{-1}X_-u\label{conu}\eeqa yields as its zero-curvature condition
\beq \tilde d\tilde A +\tilde A\wedge\tilde A=0\label{zcc}\eeq
the Toda equation \beq \partial_-\left(h^{-1}\partial_+h\right)-
\left[X_+,h^{-1}X_-h^{-1}\right] =0.\label{eqmn}\eeq Of course this
connection is not unique, any gauge-transformed version will also
yield \ref{eqmn} as its zero-curvature condition. Often it is useful
to transform with either $u$ or $u^{-1}$ yielding \beqa A_+ & = &
hX_+h^{-1}\nonumber\\ A_- & =& -\partial_-hh^{-1} +X_-\label{conhm}\eeqa or
\beqa A_+ & = & h^{-1}\partial_+h+X_+\nonumber\\ A_- & =&
h^{-1}X_-h\label{conhp}.\eeqa

This presentation of Toda systems as a zero-curvature condition is
very useful from the point of view of extracting the general
solution, and thus demonstrating the integrability (at least in that
sense of the word). This demonstration is quite complicated, but in the
end yields very simple and elegant results.

\subsection{General Solution}

First let us define $T\in\Gamma$ such that \beq \tilde
A=T^{-1}\tilde d T.\label{mon}\eeq Thus $T$ is just the inverse of
the classical monodromy associated with the linear problem defined
by a curvature-free connection. We define the subalgebras
$\gamma_\pm$ to be the direct sums of the graded subspaces with
respectively positive and negative grades. (When $\Gamma$ is a
simple finite-dimensional Lie algebra these subalgebras will be
nilpotent.) From these definitions we get the subgroups $\Gamma_\pm$
by exponentiation. We introduce \cite{LS83} the
modified Gaussian decompositions of $T$ \beq
T=M_+N_-g_+=M_-N_+g_-,\label{mgd}\eeq where $M_\pm\in\Gamma_\pm$,
likewise for $N_\pm$, and $g_\pm\in\Gamma_0$. The existence of this
decomposition is not proven in \cite{LS83}, but it seems likely
that it will be valid except on the union of some submanifolds of
$\Gamma$ of dimensions less than that of $\Gamma$.

Now suppose that we impose\beq \partial_\pm
M_\pm(x^\pm)=M_\pm(x^\pm) L_\pm(x^\pm),\label{firord}\eeq where
$L_\pm$ are some arbitrary elements of $\Gamma_{\pm 1}$. The bulk of
the derivation  of the general solution
consists of showing that this choice is actually
sufficient to reproduce the most general form for the connection
$\tilde A$, given the definitions \ref{conu}, \ref{conhm}, \ref{conhp} above.
The
method of proof begins by deriving some relations which the various
components in the decomposition \ref{mgd} must satisfy on account of the two
equivalent possibilities. Re-write this equivalence in the
form \beq R=M_+^{-1}M_-\equiv N_-gN_+^{-1},\label{defr}\eeq where
$g=g_+g_-^{-1}$. The
group element $R$ thus defined satisfies the equations of motion
\beqa \partial_+R &=&-L_+R,\label{rp}\\ \partial_-
R&=&RL_-,\label{rm} \eeqa using the definition of $R$ purely in
terms of $M_\pm$. Now substitute in the equivalent definition in
terms of $N_\pm$ and $g$. From \ref{rp} we obtain \beq
N_-\partial_+gN_+^{-1}+ \partial_+N_-gN_+^{-1}+N_-g\partial_+
N_+^{-1} =-L_+N_-gN_+^{-1}\eeq which can be re-written in the form
\beq g^{-1}\partial_+g+\partial_+N_+^{-1}N_+g^{-1}N_-
\partial_+N_-g= -g^{-1}N_-^{-1}L_+N_-g.\label{stage}\eeq We can now
use the grading structure in $\gamma$ to decompose this expression into two
separate equations. The rhs of \ref{stage} contains only grades $+1$
and below. We can separate the components of strictly positive grade
(i.e. +1) on both sides
to obtain \beq \partial_+N_+^{-1}N_+=-g^{-1}L_+g.\label{npmn}\eeq
Subtracting this off \ref{stage} yields \beq\partial_+gg^{-1} +
N_-^{-1}\partial_+N_-= -N_-^{-1}L_+N_-+L_+.\label{rest}\eeq An
exactly analogous procedure for equation \ref{rm} gives \beq
gL_-g^{-1} =N_-\partial_-N_-\label{nmmn}\eeq and \beq
g^{-1}\partial_-g +\partial_-N_+^{-1}N_+=
N_+^{-1}L_-N_+-L_-\label{rest2}\eeq

Substitution of \ref{mgd} into \ref{mon} results in an expression
for $\tilde A$ which we can be reduced using \ref{npmn}-- \ref{rest2} to
give the following simple expressions for $A_\pm$: \beqa A_+&=&
g_-^{-1}\partial_+g_- +g_+^{-1}L_+g_+,\label{ap}\\ A_-&=& g_-^{-1}
L_- g_- +g_+^{-1}\partial_-g_+\label{am}\eeqa This is most of the
work that we need to deduce the general solution. Comparing this
result with an expression for the connection in terms of the
dynamical variable $h$, for example \ref{conhp}, we find that the
choices \beqa L_\pm&=&r_\pm^{\pm 1}(x^\pm)X_\pm r_\pm^{\mp 1}(x^\pm),
\label{ldef}\\r_+&=&g_+,\label{gpdef}\\ h&=&r_-g_-,\label{hdef}\eeqa
reproduce \ref{conhp} when substituted into \ref{ap}, \ref{am}. Here
$r_\pm(x^\pm)$ are arbitrary chiral fields. This
establishes the correctness of the choice \ref{firord}. From
\ref{hdef}, \ref{gpdef} and \ref{defr} we deduce the solution for
$h$ in the form \beq
h=r_-g^{-1}r_+ =r_-N_+^{-1}M_-^{-1}M_+N_-r_+.\label{hsol1}\eeq

Of course this insufficient to explicitly evaluate a solution for $h$
as the fields $N_\pm$ are still undetermined. Leznov and Saveliev
have found an elegant solution to this problem using the
representation theory of $\Gamma$ and $\gamma$. There will be
representations of $\gamma$ with special vectors which are
annihilated by all the elements of $\gamma_+$, and in their duals
there will be vectors which are annihilated by $\gamma_-$ on the
right. Let us denote the sets of these vectors by $I_L$ and $I_R$
respectively. Taking matrix elements of $h^{-1}$ we find
\beq\langle\chi|h^{-1} |\psi\rangle=
\langle\chi|r_+^{-1}M_+^{-1}M_-r_-^{-1}
|\psi\rangle,\label{lssol}\eeq where $\langle\chi|\in I_R$ and $
|\psi\rangle\in I_L$. This expression is only useful when
$\langle\chi|$ and $ |\psi\rangle$ belong to the same
representation, of course. By this means we have eliminated the unknown
$N_\pm$ factors at the expense of only being able to determine these
matrix elements of $h^{-1}$. By taking all possible
$\langle\chi|\qquad |\psi\rangle$ we should, in principle, be able
to reconstruct $h^{-1}$.

In the simplest examples, the abelian Toda theories which are
usually studied, the gradation of $\gamma$ is just by the height of
the roots. The highest-weight state of any representation will be
annihilated by all of the step-operators corresponding to positive
roots. Since in this case $\gamma_0$ is just the Cartan subalgebra,
$h^{-1}$ lies in the maximal torus of $\Gamma$ and so we can
reconstruct it by taking the diagonal matrix elements between
highest-weight states of only the fundamental representations of
$\gamma$. This result was used extensively in \cite{OTU93} and
\cite{OU93} to discuss the soliton solutions.

The purpose of establishing the above results is that the general
solution can often be used to simplify expressions which we might wish to
calculate. These properties will be used extensively in the rest of
this paper.

\section{Defining the Non-abelian Affine Theories}

\subsection{Gradations and Embeddings}

The last section shows that, in order to define an integrable theory of Toda
type, we need an integer-graded Lie algebra and some specified
elements $X_\pm\in \fing_{\pm 1}$. Then the equation of motion
\ref{eqmn} with the field $h$ lying in $\Gamma_0=\exp(\gamma_0)$ will always
be integrable. The question then arises as to what further
conditions we should impose on $X_\pm$ in order that \ref{eqmn} be
of physical interest. There are a variety of approaches to this
problem. In the original work of \cite{LS83} both affine and finite
non-abelian Toda theories were discussed. The essence of their
approach is the following: we consider the inequivalent embeddings
of $A_1$ subalgebras (with Chevalley basis $H$, $T_\pm$) in the finite algebra
$\fing$\footnote{From now on we will use $\fing$ when $\gamma$ is a
finite algebra and $\gh$ when it is a Kac-Moody algebra.}. These were
classified by Dynkin \cite{Dy57}. Dynkin proves that the expansion
of $H$ over the generators $\{h_i\}$ of a Cartan subalgebra of
$\fing$ is sufficient to
uniquely determine the embedding, up to equivalence. The adjoint
action of the embedded subalgebra can then be used to split $\fing$
into a set of submodules, each labelled by the spin $l$. Clearly the
adjoint action of $\frac{1}{2}H$ provides a half-integer grading of
the algebra, thus

\beq \left[\frac{1}{2}H,\fing_M\right]=M\fing_M\label{grad}\eeq
and

\beq \fing=\bigoplus_{M=-L}^{L}\fing_M\eeq
where $L$ denotes the greatest value of $l$.

When all the $l$ are integers,
the embedding is said to be integral, giving a gradation of $\fing$
over the integers. The embedding also singles out a natural choice
of the elements $X_\pm$, namely $T_\pm\in\fing_{\pm 1}$. In this
case \ref{eqmn} is often called conformal Toda, of either abelian or
non-abelian type. Note that the field $h$ will
generally {\bf not} lie in an abelian group.

The much-studied abelian Toda models correspond to one particular
embedding, called the principal embedding. In this case $H$ has the
property that $\alpha_i(H)=2$ for all the simple roots $\alpha_i$.
The principal gradation given by the adjoint action of
$\frac{1}{2}H$ is the usual gradation into step-operator subspaces
of equal root height. In particular $\fing_0$ is just the Cartan
subalgebra. The spectrum of $l$ in this special case runs over a set
of integers called the exponents of the Lie algebra and $L$ is one
less than the Coxeter number \cite{Ko59}.  These abelian systems are
well-represented in the literature for their wealth of structure, in
particular for providing concrete example of W-algebras. (For a
review of this vast subject see \cite{BS92}. W-algebra structure of
the more general non-abelian theories is discussed at the classical
level in \cite{ORTW92}.)

All these types of theory based on finite $\fing$ can also be
obtained by gauging the WZNW model (see \cite{FWBFO89}), which
provides another way to obtain the Leznov-Saveliev solution
\ref{lssol}.

In what remains of this paper we shall be interested in the affine
Toda theories. In \cite{LS83} the authors proceed to the affine
non-abelian Toda theory via the above considerations. Having defined
a gradation of $\fing$ using an $A_1$ embedding they consider
$\fing$ as a subalgebra of the
corresponding loop algebra in the usual way, i.e.
$\fing\longrightarrow\fing\otimes{\Bbb C}[t,t^{-1}]$. They define a
gradation in the loop algebra using the adjoint action of
$\frac{1}{2}H+(L+1)d$ ($d\equiv d/dt$), and define elements
$\hat X_\pm=X_\pm+t^{\pm 1}X_{\mp M}$, where $X_{\mp M}$ are some
arbitrary elements of $\fing_{\mp L}$; the $\hat X_\pm$ are then used to define
the theory via \ref{eqmn}. This presentation has the advantage that
each affine theory appears as a deformation of the corresponding
finite theory. Unfortunately it has the disadvantage that the
definition is too vague to allow much progress to be made. I try to
explain how a related definition will have physical as well as
calculational advantages.

\subsection{Abelian Toda as a Model Toda Theory} The key to this
alternative definition is the observation in \cite{OU93} that, in the
special case of the abelian theory, the elements $\hat X_\pm$ can be
related to the lowest elements of a graded Heisenberg subalgebra of
the Kac-Moody algebra. From the above definitions we would obtain
expressions

\beq \hat X_\pm=\sum_{i=1}^r \sqrt k_i X_{\pm\alpha_i}+t^{\pm 1}
X_{\mp\psi}\eeq
where $k_i=\sum_j K_{ij}^{-1}$. The point is that in this case the coefficients
of the step-operators can be changed by rescaling the coordinates
and shifting the origin of the fields, without affecting the
integrability of the theory. In \cite{OTU93} the choice

\beq \hat X_\pm=\sum_{i=0}^r \sqrt m_i\hat X_{\pm \alpha_i}\eeq was
made\footnote{ Actually to make the equation of motion look more
like those generally studied in elementary particle physics we
generally introduce some extra factors in the definition of $\hat
X_\pm$-- a mass scale and a coupling constant. These do not affect
the integrability, of course.}. The advantage of this choice is that
the commutation relation

\beq \left[\hxp,\hxm\right]=K \label{hber1}\eeq
is satisfied. Working in the unextended loop algebra (or
equivalently a realisation of the Kac-Moody algebra in which $K=0$)
this means that $\hxp$ and $\hxm$ commute, and so by comparing with
\ref{eqmn} we see that the theory admits solutions with  $h$
constant and commuting with $\hxm$. In this abelian case this means
that $h$ has to lie in the
centre of $G$. The existence of well-defined degenerate vacuum
solutions means that we would expect to find soliton solutions as
the solutions of minimal energy which interpolate them. It is also
the origin of the so-called affine Toda particles, which are the
quanta of the modes obtained by diagonalising the quadratic term in
the potential about this minimum, as usual in a perturbative quantum
field theory. These particles have been extensively studied in the
literature; their masses have an interesting algebraic origin, and
an exact S-matrix has been postulated. (For an up-to-date account
see e.g. \cite{FLO91}, \cite{Do92}, \cite{DGZ92}).

\subsection{Kac-Peterson Construction}
The problem, then, is to generalise this procedure to the rest of
the affine Toda theories. Recall that we are searching for a
definition of $\hat X_\pm\in\gh_{\pm 1}$ such that the commutation relation
\ref{hber1}
will still be valid. To do this we describe an alternative
presentation of the Kac-Moody algebra due to Kac-Peterson \cite{KP85}.

Given a finite simple Lie algebra $\fing$, let us fix a Cartan
subalgebra $H_0$, and a set of simple roots $\alpha_i$ with respect
to it. Consider an element $w$ belonging to some particular
conjugacy class of the Weyl group. It is well known \cite{Hu72} that
$w$ may be lifted to an inner automorphism $\sigma_w$, implemented by
conjugation with $S_w$, of $\fing$
which acts as $w$ on $H_0$, which is canonically isomorphic to the
root space $H^\ast_0$. This automorphism is of not unique
since any redefinition of $S_w$ of the form $S_w\longrightarrow
T_1S_wT_2$, where $T_1$, $T_2$ are elements of the maximal torus of
$\fing$, ie. they commute with of all of $H_0$, will also have the
right action on $H_0$. Let us
denote the order of $w$ by $m$. Then $S_w$ splits $\fing$ into a
direct sum of eigenspaces of eigenvalues $\kappa=e^{2\pi ik}$, where
$k\in{\Bbb Z}/2m$. (Note that $(S_w)^m$ is not necessarily the
identity, this occurs for much the same reason that the spinor
representation of SO(3) is double-valued.) When all of the $k$ lie
in ${\Bbb Z}/m$ we say that the class is integral. We denote the eigenspace
of eigenvalue $\kappa$ by $\fing_\kappa$. $S_w$ can be written in the form

\beq S_w=\exp\left(2\pi i x\right).\eeq
Imposing the additional requirements that $(x,H_0)=0$, and
$[x,\fing_0]=0$ we are now in a position to define a related basis
for the loop algebra $\tilde \fing=\fing\otimes {\Bbb C}[t,t^{-1}]$. We write

\beq a'(k)=t^kt^{-x}a_\kappa t^{x} \label{basdef}\eeq
where $a_\kappa\in\fing_\kappa$ is the projection of $a\in \fing$
onto the subspace $\fing_\kappa$. Note that this is an element of
$\tilde\fing$, as can be checked by taking $t\longrightarrow e^{2\pi
i}t$. $a'(k)$ can in turn be lifted to the full Kac-Moody algebra
$\fing\otimes {\Bbb C}[t,t^{-1}]\oplus {\Bbb C}d\oplus{\Bbb C}K$
using the formula

\beq a(k)=a'(k)-\delta_{k,0}(x,a)K\label{defapr}\eeq

The algebra is ${\Bbb Z}/2$-graded using the adjoint action of
$d'=m(d+x)$, i.e. \beq [d',a(k)]=mka(k);\label{gradpr} \eeq as we
shall later show, this $d'$ is not necessarily the same as the one defined
earlier
for some appropriate choice of embedding. Note that \ref{defapr} is
equivalent to the condition that $(d',a(k))=0$. From the definition
\ref{defapr} we get the following commutation relation, which is valid provided
that either $a$ or $b$ is a pure eigenvector of the conjugation by $S_w$:

\beq \left[a(k),b(l)\right]= [a,b](k+l)+K\delta_{k+l,0}(a,b)k.\label{comr} \eeq
The last term in this expression is most easily deduced by taking the inner
product of both sides of the equation with $d'$, and using the
invariance of the Killing form. Recalling that $H_0$
is invariant under $\sigma_w$ we can decompose it into a direct sum
of graded eigenspaces, and choose a basis
$h_\kappa^\tau\in\fing_\kappa$ such that

\beq (h_{\kappa_1}^{\tau_1},h_{\kappa_2}^{\tau_2})
=m\delta_{\tau_1,\tau_2}\delta_{1,\kappa_1\kappa_2}. \eeq
The index $\tau$ allows for possible degeneracy of the spectrum.
With this convention and the definition $\hat E^\tau_{mk}\equiv
h^\tau_\kappa(k)$ \ref{comr} yields

\beq
\left[\hat E^\tau_I,\hat E^\upsilon_J\right]=
\delta_{IJ}\delta_{\tau\upsilon}IK.\label{hsal} \eeq
This is a graded Heisenberg subalgebra of $\gh$. At the end of this
section we will also detail a suitable basis for the rest of the
algebra, although this is not necessary to define the theory we will
need it later.

We repeat the fact that the grade $\pm 1$ elements $\hxp$ and $\hxm$
commute, up to an element of the centre, was crucial for the
existence of a non-trivial vacuum structure, and hence soliton
solutions, of abelian affine Toda theory. We now explain a means to
achieve this structure in the more general case.

\subsection{Conjugacy Classes and $A_1$ Embeddings}
\label{ccem}
The question we have to answer is this: under what conditions will
the graded Heisenberg algebras which are automatically produced by
the Kac-Peterson procedure contain elements with grades $\pm 1$. The
possible conjugacy classes of the Weyl groups were discussed by
Carter \cite{Ca72}. Unfortunately the picture that emerges is
clouded by a number of exceptional cases, but the vast majority of
the classes are easy to describe.

Firstly, we define a regular semisimple subalgebra of $\fing$ to be
a semisimple subalgebra which is a $\Bbb
C$-span of some subset of the step operators. Such subalgebras were
classified by Dynkin \cite{Dy57}, as a step in the classification of
non-conjugate embeddings of $A_1$ into $\fing$. Aside from a few
exceptions, the $A_1$ embeddings are principal in some regular
semisimple subalgebra (RSS). Notice that, as a consequence of the
definition, the root system of this subalgebra is just a subsystem
of that of $\fing$. Carter's result is that the majority of classes
of the Weyl group of $\fing$ contain elements which are in the Coxeter
class of some RSS. In the case of the algebras of types $A_r$,
$B_r$, $C_r$ and $G_2$ all of the classes are of this form.

Before turning to the exceptional cases we discuss the implications
for the construction of affine Toda theories of this result. Clearly,
 to have a grade 1 element\footnote{The non-degeneracy  of the
Killing form on the Heisenberg subalgebra means that the existence
of a grade 1 element guarantees the existence of a grade -1 element}
we must have an eigenvalue of the action of $w$ on the root space
which is the $e^{2\pi i/m}$. We investigate under what conditions
this will be the case.

 Let us fix the Cartan subalgebra $H_0$ of $\fing$. Fix also
a regular semisimple subalgebra $\fing_R$ of $\fing$, and thus the
element $w$, and the automorphism ${\rm Ad}S_w$. $\fing_R$ has a Cartan
subalgebra $H_R\subset H_0$ \cite{Dy57}. We can orthogonally
decompose with respect to the Killing form, $H_0=H_R\oplus H_\bot$,
where $(H_R,H_\bot)=0$. Because $H_\bot$ is thus orthogonal to all
of the simple roots of $\fing_R$ it commutes with the whole of it.
This fact is of interest in the discussion of some the physical
structure of the non-abelian theories (appendix \ref{constants}).

Now, the action of the Coxeter element is well known (originally
from the work of \cite{Ko59}. In particular the root s splits into
eigenvectors of eigenvalues $\kappa=e^{2\pi ik}$ where $mk$ runs
over the exponents of the algebra, {\em and $1$ is always a member
of this set}. When $w$ belongs to the Coxeter class of some
$\fing_R\subset\fing$ the position is not that different; $H_R$ will
decompose into subspaces according to the exponents of $\fing_R$.
Because $H_\bot$ commutes with all of $\fing_R$ it commutes with
$S_w$ and so its eigenvalue under conjugation will simply be unity.
When $\fing_R$ is  actually simple this means that there is always a
unique eigenvalue $e^{2\pi i/m}$, and so there is a natural and
unique choice of $\hxp$ as the corresponding grade 1 element of the
Heisenberg subalgebra. Now when $g_R$ is not simple we know that it
decomposes into a direct sum of simple ideals. The order of the
Coxeter element of $g_R$ is the least common multiple of the orders
of the Coxeter elements of these ideals, which are just their
Coxeter numbers. The only circumstance in which $w$ can have the
appropriate eigenvalue, and thus allow us to define $\hxp$, is when
this least common multiple is actually equal to the greatest of
these Coxeter numbers (of which all the others must be factors). If
there is more than one ideal with this Coxeter number then the
eigenvalue will be similarly degenerate. Corresponding to such $g_R$
will a be whole class of non-abelian Toda theories.

The relationship of this definition to that already discussed
\cite{LS83} is now clearer. Instead of adding $t^{\pm 1}X_{\mp M}$
to $X_\pm$  to produce elements $\hat X_\pm$ of appropriate grade, we
define a different gradation of the Kac-Moody algebra, using
$d'=m(d+x)$ in place of $\frac{1}{2}H+(L+1)d$, and append $t^{\pm
1}X_{\mp M_R}$, where $M_R$ denotes the maximal root {\it of the
embedded semisimple subalgebra $\fing_R$}, as opposed to the maximal
root(s) of the whole algebra. Only when $M_R$ is one of these will
the new definition of the theories be consistent with the previous
one.

The next task is to deal with the remaining, exceptional classes of
the Weyl group. These are classified in \cite{Ca72}, to which we
refer for further details as no particular picture emerges.
\cite{Ca72} also contains a table of the characteristic polynomials  of such
classes. Some of them do indeed have roots  $e^{2\pi i/m}$ and so
would seem to be suitable for defining exceptional Toda theories.

This completes our discussion of the possibilities for defining Toda
theories. From now on we assume that we have a chosen one of  the
suitable $w$ and specified the elements $\hat X_\pm$. We have already
remarked that the admissibility of soliton solutions is guaranteed
with the new approach, and we shall make that more concrete later
on.

\subsection{$\faz$ and Vertex Operators}
So far we have only made use of the Kac-Peterson basis for the
purpose of defining the theory, and to do this we only needed the
affinisations of the elements of the Cartan subalgebra of $\fing$.
To actually use the Leznov-Saveliev method of solution we will need
a basis for the rest of $\gh$ which we now discuss.

A logical basis to try would be that spanned by elements of the form
$X_\alpha(k)$, i.e. by affinising the eigenvector components of the
step-operators of $\fing$.  Let us denote the
number of roots on the $w$-orbit of $\alpha$ by $m_\alpha$. We can use the
formula \beq
a_\kappa=\frac{1}{2m_\alpha} \sum_{p=0}^{2m_\alpha-1}
\kappa^{-p}\sigma^p_w(a),\label{dec}\eeq  where $\kappa^{2m_\alpha}=1$ to
explicitly decompose an
element $a\in\fing$. When applied to $X_\alpha$ we see that each of
the graded components must be a linear combination of the
step-operators $X_\beta$, where $\beta$ lies on the same
$w$-orbit of the root system as $\alpha$. From the expression
\ref{dec} we can see that \beq (X_{\sigma^q(\alpha)})_\kappa=
\kappa^q(X_{\alpha})\kappa.\label{lindep}\eeq This means that to get
a proper basis we only need to use the graded components of {\em
one} representative
of each of the orbits. Furthermore, since there are
$m_\alpha$ linearly independent step operators on a
$\sigma_w$-orbit, we conclude that there are precisely $m_\alpha$
non-zero $(X_\alpha)_\kappa$. There are just two possibilities for
what the eigenvalues $\kappa$ can be, depending on the the
properties of $\sigma_w$. In the first case
$\sigma^{m_\alpha}(X_\alpha)=X_\alpha$. Then we can see by
inspection of \ref{dec} that only the $(X_\alpha)_\kappa$ for which
$\kappa^{m_\alpha}=1$ can be non-zero. In the second case,
$\sigma^{m_\alpha}(X_\alpha)=-X_\alpha$; this time
$\kappa^{m_\alpha}=-1$ is the appropriate condition. We stress here
that a class may be integral, and yet still have orbits of order
$m_\alpha$ upon whose step operators $\sigma^{m_\alpha}_w=-1$.

{}From now on we shall denote the quantities $X_\alpha(k)$ by $\hat
F_{(km)}(\alpha)$. From \ref{comr} we deduce \beq
\left[\hat E_I^\tau,\hat F_J(\alpha)
\right]=\alpha(h^\tau_\kappa)\hat F_{I+J}(\alpha),\label{comef}\eeq with
$\kappa=e^{2\pi i I/M}$.

Kac-Peterson then show that, in the basic representation of the
simply-laced affine algebras, the $\hat F_J(\alpha)$ can be realised
as the modes of a vertex operator, constructed out of the $\hat
E_I^\tau$. We shall only describe this construction briefly,
extracting the features that are of importance in the rest of our
argument. We  define the quantities \footnote{When the class of $w$
is integral we will only have integral powers of $z$ appearing in
\ref{ver}.} \beq \faz=\sum_k z^{-mk}\hat
F_{mk}(\alpha).\label{ver}\eeq Note that \beq \hat
F(w(\alpha),z)=\pm\hat F(\alpha,e^{2\pi i/m}z), \label{forbits}\eeq
where the sign depends on how we choose the step operator basis for
$\fing$. The commutation relation \beq
\left[\hat E_I^\tau,\faz
\right]=\alpha(h^\tau_\kappa)z^I\faz\label{comver}\eeq follows
directly from the equation \ref{comef}. From this it is
straightforward to show  that $U_\alpha(z)$, given by \beq
U_\alpha(z)= \exp\left(\sum_{k<0}\frac{1}{k}z^{-mk}
h_\alpha(k)\right) \faz \exp\left(\sum_{k>0}\frac{1}{k}
z^{-mk}h_\alpha(k)\right). \label{deft}\eeq has the property that it
commutes with the Heisenberg subalgebra spanned by the $\hat
E^\tau_I$. It can be shown that the various $U_\alpha(z)$ form an
abelian {\em group}, extended by the complex numbers. It turns that
we have the vertex operator representation  \beq \faz=\colon
\exp\left(-\sum_{k\neq 0}\frac{1}{k}z^{-mk} h_\alpha(k)\right)\colon
U_\alpha(z)\label{verrep}\eeq where the colons denote normal
ordering, familiar from quantum field theory. The space upon which
$\faz$ acts is the tensor product of a certain representation of the
extended group, and the Fock space built up by the $E_I^\tau$, for
which $I<0$. $U_\alpha(z)$ can be thought of as taking account of
the zero modes, c.f. the homogeneous vertex operator construction
discussed in \cite{GO86}, for example.

The only element of this construction that we will make use of
concerns the expression for the normal-ordered product of two of the
$\faz$. Using an argument essentially the same as that in
\cite{OU93} we find \beqa \lefteqn{\quad\colon \hat
F(\alpha,z_1)F(\beta,z_2)\colon=S_{\alpha\beta}(z_1,z_2) \colon
\exp\left(-\sum_{k}\frac{1}{k}\left[ z_1^{-mk} h_\alpha(k)+
z_2^{-mk} h_\alpha(k) \right]\right)\colon}\nonumber\\
&\hspace{6cm}\times\colon U_\alpha(z_1)U_\beta(z_2)\colon.&\hspace{3cm}
\label{normord}\eeqa
The numerical `S-matrix' factor $S_{\alpha\beta}$ is given by \beq
S_{\alpha\beta}(z_1,z_2) =\prod_{n=0}^{m-1} \left(1-e^{-2\pi
in/m}\frac{z_2}{z_1} \right)^{w^n(\alpha)\cdot\beta}\label{smat}\eeq
{}From this we can deduce the key fact that $(\faz)^2=0$. The reason
this is important is that it means that $\faz$ is nilpotent within
{\em} any unitary highest weight representation, and so the `group element'
$\exp(\faz)$ has a well-defined meaning.

In conclusion, it seems clear that this definition of the general
affine Toda theory has at least one advantage over the traditional
one, in that we now have a concrete presentation related from the
outset to vertex operators.
\section{Abelianisation}

Having established a straightforward description of the Toda
theories of affine type we now consider how to
extract the simplest conserved currents using the abelianisation
procedure (see \cite{OT2} and references therein).

\subsection{Review}

We include a description of the abelianisation procedure in the
spirit of \cite{OT2}, but substantially revised to suit the greater
generality of the systems considered here. The essential principle
is the following, suppose that we can find some gauge transformation
of the connection \ref{conhp} which will leave it lying in some
subalgebra of $\gh$ commuting with a particular element $X$, the
centraliser $Z(X)$ of $X$. Such a gauge transformation will of
course depend on the dynamical variables, just as the transformation
linking \ref{conu} with \ref{conhm} and \ref{conhp}  did. Suppose
$Y$ is some element of the centre of $Z(X)$\footnote{The elements
$Y$ are discussed in appendix \ref{centre}}: taking the inner
product of $Y$ with the zero-curvature condition \ref{zcc}, the
commutator term drops out and so we are left with \beq \tilde
d\left( \tilde A',Y\right)=0.\label{cons}\eeq Note that we have
denoted the transformed connection by $\tilde A'$. This can be rewritten as
the conservation equation \beq\delta(\ast\tilde A',Y)=0.\label{cA}\eeq
{}From this conserved current we can of course extract a conserved
charge in the usual fashion.

In the case of the affine Toda theories the grading structure gives
us a means of actually implementing the abelianisation procedure. Starting with
the connection \ref{conhp} we note that the only component with a
positive grade is the $\hat X_+$. This raises the possibility that
we might be able to use gauge transformations in $\hat G_-$, which
would leave the positive-graded component invariant, since $(f\hat
X_+ f^{-1})\cap\gh_+=\hat X_+\forall f\in\Gamma_-$. Now we use the
following iterative procedure. Although such a gauge transformation
must necessarily depend on an infinite number of parameters, it can
in fact be described iteratively, using the following lemma:

\begin{lemma} Suppose that we have managed to perform a
gauge transformation such that the only non-zero components of the
connection which do not commute with $\hxp$ are in grades $n$ and
below. Then there exists a gauge transformation by a $\hat G_-$
valued section of the form $\exp(\omega_{n-1})$, with
$\omega_{n-1}\in\gh_{n-1}$ which will cancel the component of
$\tilde A$ in $\gh_n$ which does not commute with $\hxp$.\end{lemma}

The proof of this is quite straightforward, given the new definition
of the non-abelian affine Toda theory. A transformation of this
form will change the connection by adding
$[\omega_{n-1},\hxp]\in\gh_n$, plus some terms of lower grade which
are not of interest here. Thus all we have to prove is that $\gh_n$
is the direct sum of its intersection with $Z(\hxp)$ and
$\ad(\hxp)\gh_{n-1}$. This proof is straightforward if we use the
results of the previous section. We established that there was a
graded basis for the Kac-Moody algebra, such that $\gh_n$ is spanned
by the $\hat E^\tau_n$ and the orbit representatives $\hat F_n(\alpha)$.
Using the commutation relation \ref{comef} and the fact that $\hxp$
can be expanded over the $\hat E^\tau_1$, we see that the only elements
of this basis for $\gh_n$ which do not commute with $\hxp$ must
indeed be the ad$\hxp$ images of elements of the basis of $\gh_{n-1}$.

It is clear from this result that the desired transformation can be
found iteratively by removing the non-commuting component of
successively lower graded subspaces, starting with $\gh_0$. By the
use of an exactly analogous argument starting with the connection
\ref{conhm} and proceeding in the opposite direction we will find
another infinite set of conserved currents.

The gauge transformations are not unique, since any transformation using
an element of the exponential of $Z(\hxp)$ will leave the connection
in $Z(\hxp)$. We now exploit this freedom to find a particularly nice
form for the lowest conserved currents, namely the chiral components
of the energy-momentum tensor.

\subsection{Explicit Expression for the Lowest Currents}

{}From now on we restrict attention to the theories for which $w$ belongs
to an integral class (but see \cite{GORS93} for a partial discussion of
the general case).
In order to find explicit expressions for the currents we need to
know the elements of the centre of $Z(\hxp)$. A form for these is
conjectured in appendix \ref{centre}, but here we shall only need the canonical
element, namely $\hxp$ itself. To calculate its inner product with
the transformed connection we need only know the component of this
connection lying within $\gh_{-1}$. This we can calculate using only
the first two steps of the iterative procedure.

Starting from the form for the connection \ref{conhp} we derive the
condition for $\omega_{-1}$, finding \beq
\left[\hxp,\left[\omega_{-1},\hxp\right]+h^{-1}\partial_+h\right]=0.
\label{abscon} \eeq  Performing the next transformation as well we
find the component of $A_+$ with grade -1 to be \beq
[\omega_{-2},\hxp ] +\frac{1}{2} [\omega_{-1},[\omega_{-1},\hxp
]]+[\omega_{-1},h^{-1}\partial_+h] -\partial \omega_{-1}].\eeq
Taking the inner product with $\hxp$ and using the invariance of the
Killing form we get \beq (\hxp,A_+)=\frac{1}{2}
(h^{-1}\partial_+h,[\hxp,\omega_{-1}])-
\partial_+(d',[\hxp,\omega_{-1}])\label{chiral}\eeq The $d'$ has
been introduced purely for convenience. Now in general the solution
of \ref{abscon} for $[\hxp,\omega_{-1}]$ will be $h^{-1}\partial_+h$
plus a term in the kernel of $\ad \hxp$. Denote by $P$ the
orthogonal projection into this kernel, then
$[\hxp,\omega_{-1}]=(1-P)h^{-1}\partial_+h$. Substituting into
\ref{chiral} yields  \beq (\hxp,A_+)=\frac{1}{2}
(h^{-1}\partial_+h,h^{-1}\partial_+h)-
\partial_+(d',h^{-1}\partial_+h)+\ldots,\label{chiralt}\eeq where
the dots denote some terms which can be seen to be independent of
$x^-$ by applying the projection $P$ to the equation of motion (or
alternatively dispensed with by exploiting the remaining gauge
freedom). These can be dropped (we discuss the field
$Ph^{-1}\partial_+h$ in appendix \ref{constants}). We can easily
show that the transformed $(\hxp,A_-)$ vanishes due to the equation
of motion, (aside from more terms similar to the above) so this
means that we have a chiral conserved current given by
$J^-=(\hxp,A_+)$, $J^+=0$. An analogous expression for another
chiral current can be derived by starting from \ref{conhm} and
performing gauge transformations with elements of $\hat G_+$.

We shall now show that we can use the results needed for the
derivation of the general solution to greatly simplify the
expression \ref{chiralt}.

\subsection{Chiral Current in Terms of Free Fields}

Using the expression for $h$ given by \ref{hsol1} we can deduce the
following formula for $h^{-1}\partial_+h$: \beq h^{-1}\partial_+h=
r_+^{-1}\partial_+ r_+ -r_+^{-1}\partial_+gg^{-1}r_+ \eeq We find for
the `kinetic term' of the expression for the conserved charge \beqa
\lefteqn{\frac{1}{2}(h^{-1}\partial_+h,h^{-1}\partial_+h)}
\nonumber\\ &&=\frac{1}{2} (r_+^{-1}\partial_+r_+
,r_+^{-1}\partial_+r_+) -(\partial_+r_+r_+^{-1},\partial_+gg^{-1})
-(L_+,N_-L_+N_-^{-1}).\label{first}\eeqa To arrive at this formula we
make extensive use of the grading properties of the inner product,
and \ref{rest} from the derivation of the Leznov-Saveliev solution.

The second term is trickier to reduce. We get \beq
\partial_+(d',h^{-1}\partial_+h)=\partial_+\left(d',
\left(r_+^{-1}\partial_+r_+N^{-1}_-L_+N_-\right)\right),
\label{redo}\eeq where we have made use of \ref{rest} in particular.
Expanding the derivative on the last term of this expression gives
\beqa \lefteqn{\left(d',\left[N_-^{-1}L_+N_-,N_-^{-1}\partial_+N_-\right]+
\left[+N_-^{-1} \partial_+r_+r_+^{-1}N_-,N_-^{-1}L_+
N_-\right]\right)\qquad\qquad} && \nonumber\\
&&=\left(d',\left[N_-^{-1}L_+N_-,N_-^{-1}\partial_+N_- -N_-^{-1}
\partial_+r_+r_+^{-1}N_- \right]\right). \label{redt}\eeqa Now we
examine the grading structure of this expression. Because this
contains a commutator of a quantity of grades +1 and below with a
quantity of grade zero and below, it follows that the only
contribution to the inner product with $d'$ must come from the grade
1 part of the first argument of the commutator, which is just $L_+$.
This gives us \beq\left(L_+ ,N_-^{-1}\partial_+N_- -N_-^{-1}
\partial_+r_+r_+^{-1}N_- \right).\label{redth}\eeq This in turn
leads to \beq \left(L_+ ,-N_-L_+N_-^{-1} -N_-^{-1}
\partial_+r_+r_+^{-1}N_- \right),\label{redth2}\eeq when we use
\ref{rest}. We recognise the first term as cancelling part of the
`kinetic' part of the conserved current \ref{first}. For the remainder of
\ref{redth2} we use \ref{rest} again to remove $L_+$ yielding \beq
-\left(\partial_+ gg^{-1},\partial_+r_+r_+^{-1}\right),\label{fin}
\eeq which cancels with another term from the `kinetic' part \ref{first}. Thus
we are left with the remarkable result that \beq (\hxp,A_+)=\frac{1}{2}
(r_+^{-1}\partial_+r_+,r_+^{-1}\partial_+r_+)-
\partial_+(d',r_+^{-1}\partial_+r_+).\label{chirfree}\eeq Put another
way, the conserved current takes exactly the same form in terms of
the free fields. Of course this makes it obvious that it must be
chiral and conserved. The same result holds for the chiral current
got from \ref{conhm}

In \cite{OTU93} this result was proven for the abelian theories
using \ref{lssol}  and representation theory. We see from the above
that it is not actually necessary to take the matrix elements,
indeed for the non-abelian theories it would probably prove too
difficult anyway. \cite{OTU93} is still interesting because of the
way these cancellations translate into representation theory, where
they arise due orthogonality of different highest weight states in a
tensor product decomposition.

We shall now discuss how the above result
is of value in the study of the soliton solutions of the general
theories.

\section{Canonical Energy-Momentum Tensor}

In this section I present the proof that the lowest conserved
charges actually correspond to the chiral components of the improved
energy-momentum tensor.

\subsection{General Toda Action} In order to construct a canonical
energy-momentum tensor we first need an action for the theory. This
action $S$ takes the form \beq
S=S_{\rm WZNW}+2\eta\int_{\partial B}\tilde\omega\left( (\hxp,h^{-1}\hxm
h)-(\hxp, \hxm)\right),\label{act}\eeq where the
Wess-Zumino-Novikov-Witten type action is
defined to be \beqa\lefteqn{S_{\rm WZNW}
=\eta\Biggl[\frac{1}{2}\int_{\partial B}
\tilde\omega g^{\mu\nu}\left(h^{-1}\partial_\mu h,h^{-1}\partial_\nu
h\right)}\qquad\qquad&& \nonumber\\
&& \quad+\frac{1}{3}\int_B \left(h^{-1}\tilde dh\stackrel{\wedge}{,}
h^{-1}\tilde
dh\wedge h^{-1}\tilde dh\right)\Biggr].\label{wact}\eeqa Here
$\tilde\omega$ is the usual measure form on a metric manifold, and $B$ is a
manifold whose boundary is space-time. See e.g. \cite{GO86} or \cite{Bo89} for
a
review of the WZNW model. The equation of
motion \ref{eqmn} follows from the variation of this action in the
usual way.

\subsection{Splitting of Energy-Momentum Tensor}

The canonical energy-momentum tensor is \beq T_{\mu\nu}=\frac{2}{\sqrt
{|\det g|}}\frac{\delta S}{\delta g^{\mu\nu}} \label{emdef}\eeq Thus
we get \beqa \lefteqn{T_{\mu\nu}=\eta\Biggl[
\left(h^{-1}\partial_\mu h,h^{-1}\partial_\nu h\right) -
{g_{\mu\nu}\over 2} \left(h^{-1}\partial_\alpha
h,h^{-1}\partial^\alpha h\right)}\qquad\qquad &&\nonumber\\
&\hspace{2cm}&-2g_{\mu\nu} \left(\left(\hxp,h^{-1}\hxm h\right)
-\left(\hxp,\hxm\right)\right)\Biggr].\label{em}\eeqa Following much
the same procedure to that of \cite{OTU93} we try to split this
tensor into a traceless part plus a total derivative `improvement'.
Define the total derivative term $C$ by \beq
C_{\mu\nu}=2\eta\left[\left(d',\partial_\mu\left(h^{-1} \partial_\nu
h\right)- g_{\mu\nu}
\partial_\alpha\left(h^{-1}\partial^\alpha
h\right)\right)+g_{\mu\nu}\left(\hxp,\hxm\right)\right]. \label{imp}\eeq Note
that this is symmetric; this can easily be shown by expanding the
derivative and using the fact that $d'$ commutes with all of
$\gh_0$. Now we write \beq T_{\mu\nu}=C_{\mu\nu}
+\Theta_{\mu\nu}\label{split}. \eeq Calculating the components of
$\Theta$ in a light-cone basis (we need to use the equation of
motion \ref{eqmn} to eliminate some terms) we find that it is
diagonal, with \beq\Theta_{++}=
\eta\Biggl[\left(h^{-1}\partial_+h,h^{-1}\partial_{+} h\right)
-2\left(d',\partial_+\left(h^{-1}\partial_+ h\right)\right)\Biggr],
\label{thet}\eeq with an analogous expression for $\Theta_{--}$. It
is no surprise that this should be just \ref{chiralt} multiplied by
a factor $\eta$.

Thus we find the same behaviour as in \cite{OTU93}, the \footnote{When the
class of $w$ is
integral we will only have integral powers of $z$ appearing in this
expression.}
energy-momentum tensor splitting into two parts; one a traceless
part dependent only on the free fields, and the other a total
derivative. We can use this fact to calculate expressions for the
energies and momenta of the soliton solutions of these theories.

\section{Energy and Momentum of Soliton Solutions}\label{emsol}

In the previous section we obtained an expression for the energy and
momentum of a solution of the affine Toda field theories constructed
by the Leznov-Saveliev procedure. Of particular interest for physics
are the soliton solutions, which, as we have already discussed, are
expected to interpolate the degenerate vacua of the theory. In
\cite{OSU93} the specialisation of the general solution \ref{lssol}
which yields the soliton solutions was conjectured. We repeat this
conjecture here for convenience.

\subsection{Solitonic Specialisation}
Let us discuss the physical significance of the splitting
\ref{split}. $C_{\mu\nu}$ is a total derivative so it can be explicitly
integrated to yield surface terms at infinity. We would expect
the conserved charges of solitons, which are topological objects, to
have this kind of behaviour. The other term, $\Theta_{\mu\nu}$, can
be made to vanish simply by setting the free fields $r_{\pm}$ to
zero. This rather drastic choice still leaves a large class of
non-trivial solutions, as we shall show.

We can now solve the equations of motion \ref{firord} explicitly.
Thus \beq M_\pm=M_\pm(0)\exp\left(x^\pm\hat X_\pm\right),
\label{msol}\eeq where $M_\pm(0)$ are some arbitrary group-valued
constants. Inserting this into the solutions for matrix elements
\ref{lssol} gives \beq\langle\chi|h^{-1} |\psi\rangle=
\langle\chi|\exp\left(-x^+\hxp\right) {\cal G}_0\exp\left( x^-\hxm\right)
|\psi\rangle,\label{solsol}\eeq The constant ${\cal G}_0$ is equal
to $M_+(0)^{-1}M_-(0)$. In \cite{OSU93} it is conjectured that the
N-soliton solution is given by a choice of ${\cal G}_0$ of the form
\beq {\cal G}_0=\exp\left(Q_1\hat F(\gamma_1,z_1)\right)\ldots
\exp\left(Q_i\hat
F(\gamma_i,z_i)\right)\ldots \exp\left(Q_N\hat
F(\gamma_N,z_N)\right).\label{grpel} \eeq The parameters in this expression
are: \begin{itemize}

\item
$z_i$ are some complex numbers which describe the rapidities of the
component solitons, ordered so that $|z_1|\geq |z_2|\geq\ldots\geq
|z_i|\geq\ldots\geq |z_N|$.
\item
$Q_i$ describe the positions of the solitons at some initial time.
\item
$\gamma_i$ are some roots of $\fing$.  Roots belonging to the same orbit of
$w$ give the same species of soliton.
\end{itemize}
We will explain how the parameters can be given these
identifications later on.

This form for ${\cal G}_0$ has the feature that it allows
\ref{solsol} to be
considerably simplified. From \ref{comef} it is obvious that the
commutation relations \beq \left[\hat X_\pm,\faz\right]=q^\pm_\alpha
z^{\pm 1} \faz \label{defq} \eeq hold, with $q^\pm_\alpha$ easily calculable
from
our knowledge of the definitions of $\hat X_\pm$ (this is described
in appendix \ref{kpb}). Following
\cite{OTU93} we shall use conventions so that
$q^-_\alpha=(q^+_\alpha)^\ast$. (The admissibility of this
convention is a consequence of the existence of the Chevalley
involution/definition of the adjoint of an element of the Lie
algebra $\fing$.)  Using \ref{defq} we can re-write \ref{lssol} as
\beqa \lefteqn{\langle\chi|h^{-1} |\psi\rangle=\langle\chi|\exp\left(Q_1
e^{-q^+_{\gamma_1}z_1x^+- q^-_{\gamma_1}z^{-1}_1x^-}\hat
F(\gamma_1,z_1)\right)\ldots} \nonumber\\ &&\vspace{2cm}
\ldots \exp\left(Q_N
e^{-q^+_{\gamma_N}z_Nx^+- q^-_{\gamma_N}z^{-1}_Nx^-}\hat
F(\gamma_N,z_N)\right)\exp\left(-x^+x^-\frac{K(\hxp,\hxm)}{m}\right)
|\psi\rangle.\label{fulsol}
\eeqa This explains why we identify $\ln Q_i$ as describing the
relative positions of the solitons. Given this form for the soliton solutions
we now show how it
is possible to explicitly calculate the energy and momentum of such
a solution.

\subsection{Calculating Surface Terms}

Using the expression \ref{imp} for the improvement of the
energy-momentum tensor we deduce that the light cone components of
the energy and momentum of a soliton solution are of the form
\beq P^{\pm}= 2\eta\int_{-\infty}^{\infty}dx \left(\mp\partial_x
\left(d',h^{-1}\partial_\mp
h\right)+\frac{1}{\sqrt 2}\left(\hxp,\hxm\right)\right). \label{topem} \eeq
It is easy to show that the second, `classical renormalisation', term will
cancel with a
similar contribution coming from the $K$-component of equation
\ref{fulsol} substituted into the first term. This just leaves a
portion which depends on which particular soliton solution we are
considering. To evaluate this we need to introduce some elements of
representation theory.

In Appendix \ref{rhoproof} we introduce a particular highest weight
representation $V_\rhc$ of the affine algebra with the following
special properties:  \begin{itemize} \item $V_\rhc$ is a highest
weight representation with respect to the grading defined by $d'$,
i.e. all elements of positive grade annihilate the highest-weight
state $|\rhc\rangle$. \item There is a Cartan subalgebra, $H_\bigtriangleup$,
of $\gh$ contained in $\gh_0$. The orthogonal decomposition of
$\gh_0$ reads $\gh_0=H_\bigtriangleup\oplus \gh_\bigtriangleup$. $|\rhc\rangle$
is
annihilated by all of $\gh_\bigtriangleup$. \item Given $h_\bigtriangleup\in
H_\bigtriangleup$,
$h_\bigtriangleup|\rhc\rangle=\rhc(h_\bigtriangleup)|\rhc\rangle$. Thus
$|\rhc\rangle$
is a one-dimensional representation of both $\gh_0$ and $\hat G_0$.
\item The following result holds for $s\in\gh_0$: \beq
\left(d',s\right)=\langle \rhc |s|\rhc\rangle. \label{magic}\eeq
\end{itemize}
The utility of this representation is that we can
replace the inner product with $d'$ occurring in \ref{topem} with a
matrix element, which the Leznov-Saveliev solution allows us to
calculate. We have \beq P^{\pm}=\mp 2\eta\left[\left(d',h_{\rm
red}^{-1}\partial_\mp h_{\rm red}\right)\right]^{\infty}_{-\infty},
\label{topem2} \eeq where
\beqa \langle\chi|h_{\rm red}^{-1}
|\psi\rangle&=\langle\chi|\exp\left(Q_1 e^{-q^+_{\gamma_1}z_1x^+-
q^-_{\gamma_1}z^{-1}_1x^-}\hat F(\gamma_1,z_1)\right)\ldots
\nonumber\\ &\qquad \ldots \exp\left(Q_N e^{-q^+_{\gamma_N}z_Nx^+-
q^-_{\gamma_N}z^{-1}_Nx^-}\hat F(\gamma_N,z_N)\right)
|\psi\rangle,\label{fulsolred} \eeqa i.e. we have performed the
classical renormalisation. Because $|\rhc\rangle$ is a
one-dimensional representation we can re-write \ref{topem2} to read
\beq P^{\pm}= \mp 2\eta\left[\langle\rhc |h_{\rm
red}^{-1}|\rhc\rangle\langle \rhc |\partial_\mp h_{\rm
red}|\rhc\rangle\right]^{\infty}_{x=-\infty}. \label{topem3} \eeq
The second matrix element can be expressed in terms of the first,
which is given by the expression \ref{fulsolred}. Thus an explicit
calculation is feasible.  At this point it is
appropriate to discuss the reality conditions that ought to
be applied to both the definition of the theory and the solution
\ref{topem3} to produce physical soliton solutions.

\subsection{Constraints on a Physical Toda Theory}

As we remarked earlier it is customary in abelian Toda theories to
introduce some extra constant parameters into the equation of motion
\ref{eqmn}. The reason for this is that the quadratic term in the
potential\footnote{When we express $h$ as an exponential of some Lie
algebra valued field.} can be diagonalised to yield a theory of bosons, whose
interaction is governed by the higher terms. Of course this is the
procedure used in quantum field theories since their inception.  In this
approximation,
the abelian Toda equation
is classically just a number of decoupled Klein-Gordon equations
with real positive masses. \beq
\left(2\partial_+\partial_-+\mu_i^2\right)\phi_i=0. \label{kg}\eeq
This is done (e.g. \cite{OTU93}) by multiplying the definitions of $\hat X_\pm$
by $\pm \mu$, where $\mu$ is a real, positive mass parameter. We
also customarily set $\eta=1/\beta^2$, where $\beta$ is a coupling
constant whose reality properties we shall decide on later.

Let us make these changes for the more general theories that we
discuss here. We then get a revised version of equation
\ref{fulsolred}: \beqa \langle\chi|h_{\rm red}^{-1}
|\psi\rangle&=\langle\chi|\exp\left(Q_1 e^{\mu( q^+_{\gamma_1}z_1x^+-
 q^-_{\gamma_1}z^{-1}_1x^-)}\hat F(\gamma_1,z_1)\right)\ldots
\nonumber\\ &\qquad \ldots \exp\left(Q_N e^{(\mu q^+_{\gamma_N}z_Nx^+-
 q^-_{\gamma_N}z^{-1}_Nx^-)}\hat F(\gamma_N,z_N)\right)
|\psi\rangle.\label{newred} \eeqa In particular the one-soliton
solution is: \beqa \langle\chi|h_{\rm red}^{-1}
|\psi\rangle&=\sum_{n=0}^p\frac{1}{n!}\langle\chi|\left(Q \hat
F(\gamma,z)\right)^n
|\psi\rangle  e^{n\mu\left(q^+_{\gamma}zx^+-
 q^-_{\gamma}z^{-1}x^-\right)},\label{redone} \eeqa where
$p$ is the greatest non-vanishing power of $F(\gamma,z)$ within the
representation. Physically, a one-soliton solution has the property
that it can be set to rest by a real Lorentz transformation. In two
dimensions this just means changing the modulus of $z$. The time
dependence in the ansatz \ref{redone} enters through \beq
\frac{\mu}{\sqrt 2}\left(q_\gamma^+z-(q_\gamma^+)^\ast z^{-1}\right),\eeq and
so we see that such a transformation can only be performed if \beq
q^+_\gamma z=\epsilon |q_\gamma| e^\lambda,\eeq with $\lambda$ real,
and $\epsilon=\pm 1$.

The conserved charges in equation \ref{topem3} are found from the asymptotic
behaviour of the solution \ref{redone}. This is just a polynomial in \beq
\exp\left(\mu|q_\gamma| \epsilon \left(e^\lambda+e^{-\lambda}\right)x/\sqrt
2 \right),\eeq as far as the space dependence is concerned, so its
asymptotic behaviour depends on the sign of epsilon, being dominated
by the terms of greatest or least degree in the sum. Finally we obtain \beq
P^\pm =-{2\over\beta^2}e^{\mp \lambda}|q_\gamma|p_\rhc^{(\gamma)}.
\label{onechar}\eeq Note that the factor epsilon cancels out. We can
think of this as confirming that soliton and antisoliton possess the
same energy. This expression means that the energy of a
soliton with $\beta$ real would be negative, and so we require that
$\beta$ be pure imaginary on physical grounds. This behaviour is
familiar from the abelian theory \cite{OTU93}. For the invariant
masses of the solitons we find \beq M^2=2P^+P^-={8\mu^2 \over
\beta^4}|q_\gamma|^2 (p_\rhc^{(\gamma)})^2\label{invm}. \eeq

Having studied the one-soliton solution we are now in a position to
examine the reality conditions appropriate to an N-soliton
scattering solution, i.e. a solution with the group element of the
form in \ref{grpel} for which each one of the solitons can be set to
rest by an appropriate Lorentz transformation. Using an obvious
notation \beq P^\pm ={-2\mu\over
\beta^2}\sum_{i=1}^N|q_{\gamma_i}| p_\rhc^{(i)}e^{\mp \lambda_i}.
\label{multisol}\eeq This formula yields the physical interpretation
of the rest of the parameters appearing in the expression \ref{grpel}.

In addition to the N-soliton scattering state we expect, by analogy
with the abelian theories, that the spectrum will also contain
breather solutions. A breather solution is a bound state of soliton
and antisoliton whose individual energies and momenta are not real
but whose sum is. This sum can be obtained from the two-soliton
version of \ref{multisol} by writing $\lambda_1=\lambda+i\delta$,
$\lambda_2=\lambda-i\delta$ giving \beq P^\pm_{\rm
breather}={-4\mu|q_\gamma|p_\rhc \over\beta^2}e^{\pm \lambda}
\cos\delta.\eeq In fact it is possible to construct other N-soliton
solutions for whom only the total energy-momentum is   real. We
assume that these will be ruled out by reality restrictions on the
higher conserved charges. (See \cite{OU93} for a discussion of this point.)
Thus we expect the most general soliton solution which satisfies the
physical constraints to be a scattering state of solitons and
breathers which are individually physically allowed.

\subsection{Particles and Solitons}

In \cite{OTU93} a relationship between the masses of the Toda
particles and solitons in abelian Toda was described. We now want to give
arguments
which show that this is probably also the case in the more general
non-abelian theories. The difficulty here is that the equation of
motion will not in general decompose into a set of decoupled
Klein-Gordon equations for some $\gh_0$ valued fields $\ln h$. The
non-abelian nature of $\gh_0$ means that \ref{eqmn} will also
contain some terms quadratic in the first derivatives of the fields.
(In \cite{GS92} the authors show that these theories can be thought
of as integrable theories of particles moving in a black hole space-time.)
In spite of this problem we still find that the soliton solution
\ref{redone} is a polynomial in an exponential function, and the
static solution of a set of decoupled Klein Gordon equations like
\ref{kg} are just \beq \phi_i\sim e^{\pm \mu_i x}\label{kgsol}.\eeq We
might expect that as the soliton solutions asymptotically approach
the vacuum the particle modes do decouple, and so $\langle h_{\rm
red}\rangle$ ought to be a polynomial in Klein-Gordon solutions.
Using \beq M_{{\rm particle}\ \gamma}=\mu_\gamma
\hbar,\label{qmass}\eeq and comparing the dependences of \ref{redone} and
\ref{kgsol}
we find that \beq M_{{\rm soliton}\ \gamma}={M_{{\rm particle}\
\gamma}p_\rhc^{(\gamma)}\over |\beta|^2\hbar}.\label{masrel}\eeq
We stress again that this formula should not be taken too seriously
until the uncertainty in the precise meaning of $M_{{\rm pariticle}}$ is
resolved.

We shall now describe an additional property of the soliton solutions.
\subsection{Fusing Rule}

In \cite{OU93} a soliton analogue of Dorey's fusing rule for the scattering
of Toda particles was discussed. This made use of the expressions
\ref{normord}, \ref{smat} for the normal-ordered product of two of
the $\faz$ within a level one representation of a simply-laced $\gh$. By
inspection
of \ref{smat} we see that $S_{\alpha\beta}$ will be singular at
$z_2=e^{2\pi in/m}z_1$ whenever $w^n{\alpha}\cdot\beta=-1$, i.e.
when $w^n{\alpha}+\beta$ is a root of $\fing$. Suppose that we have
a two-soliton solution (with no reality condition imposed),
generated by \beq {\cal G}_0=\exp\left(Q_1\hat
F(\alpha,z_1)\right) \exp\left(Q_2\hat
F(\beta,z_2)\right).\label{fuse} \eeq Suppose that we
simulataneously take the limits \beqa z_2&\longrightarrow& e^{2\pi in/m}z_1,
\nonumber\\ Q_1,\ Q_2&\longrightarrow &0,\label{limits}\eeqa in such a way
that the normal-ordered product of the two $F$ remains
finite\footnote{We stress again that this product is understood to
be evaluated in a level one representation of a simply-laced $\gh$.
The vertex operator construction can be used to make statements
about highest weight representations of arbitrary level, as well as
representations of the non-simply laced algebras. These points are
discussed in the appendix \ref{otherrep}} By using the adjoint
action of the $\hat E_I^\tau$ we can show that the resulting ${\cal
G}$ is proportional to the exponential of a third $F$, thus: \beq  {\cal
G}_0=\exp\left(Q\hat
F(w^n(\alpha)+\beta,z_2)\right).\label{fusedf}\eeq The constant $Q$
is the constant of proportionality whose determination is irrelevant
for our purposes.

What this means physically is the following, that a two-soliton
solution can "fuse" to form a one soliton solution. The fusion is
governed by the root system of $\fing$. We know that the orbits of
$w$ correspond to the possible species of solution once we have
removed the trivial solutions (see appendix \ref{constants}). The
condition $w^n{\alpha}\cdot\beta=-1$ just means that
$w^n(\alpha)+\beta$ is a root, and so we have the following
statement of the generalised fusing rule. Take a two soliton
solution, with solitons of species $\alpha$ and $\beta$, such that
there exist elements of the $w$-orbits of $\alpha$ and $\beta$ whose
sum is also a root. The parameters of this solution can be chosen so
that in the above limit the two solitons fuse to give a third whose
species is $w^n(\alpha)+\beta$. It is important to note that if the
third soliton is physical then at least one of the originals must be
unphysical. This is probably of significance in the quantum theory
as it would seem to mean that one soliton cannot decay into two.

We
now discuss the theory.

\section{Discussion and Conclusions}

We have seen that the non-abelian affine Toda theories can be
defined in such a way that the classical soliton solutions can be
extracted in simialr fashion to that used for the solitons of the
abelian theory \cite{OTU93}. This was done by sticking to
group-valued Toda fields $h$ instead of the Lie algebra valued
fields almost exclusively used in the literature. This sort of
method is more in the spirit of the original solution (formula
\ref{lssol} of the
theories, as we have discussed. This principle has allowed us to
extend the result of \cite{OTU93} on the splitting of the
energy-momentum tensor of the abelian theories to the whole class,
which would seem computationally impossible using the means given
there.

Using results of \cite{Ca72} and \cite{Dy57} on the relationship
between conjugacy classes of the Weyl group of the
Lie algebra $\fing$ and inequivalent embeddings of $A_1$ subalgebras
we are able to give a presentation of the
theories making contact with the elegant work of \cite{KP85}, which
is essential if we are to study the soliton spectrum of the theories
in a systematic fashion. We find that there are theories in addition
to those originally discussed by Leznov and Saveliev in \cite{LS83}.

We have shown that many of the remarkable
features of the soliton spectrum of the abelian theories can be
extended to the general theories in a general fashion. In particular
we have seen how the conserved quantities are topological and real,
despite their densities being complex. Finally we note that we have
not bothered to calculate any explicit soliton solutions. There is a
very good reason for this. When the equation of motion \ref{eqmn}
for the non-abelian theories are written out in terms of some
parametrisation of the Toda field $h$ (usually in terms of some
product of exponentials of some fields lying in $\gh_0$, e.g.
\cite{LS83})  they look
extremely messy. We want to stress that this is unnecessary, since
we are able to deduce information about the theory in notation in
which it appears natural without resorting to brute-force methods.

Let us now discuss possible extensions of the work. Of course the
goal of any particle physicist is always to find a quantum theory,
where we expect even greater richness. There have been several
approaches to quantising the abelian theories which might be hoped to
work in general. In \cite{BB91}, \cite{Ba88} the authors quantise
the matrix elements of $h^{-1}$ which appear in the Leznov-Saveliev
solution. They satisfy an exchange relation which makes contact with
the theory of quantum algebras. It is possible that the present
approach will also lend itself to this means of quantisation. An
interesting alternative seems to be to try to quantise the free
fields, zero-modes and all which underly the Leznov-Saveliev
solution. This appears to be the course followed in \cite{ABBP93},
for the case of real, abelian, finite Toda theories. Such theories
and their non-abelian generalisations
have been of much interest recently on account of their having a
W-algebra structure. At the classical level there is a discussion in
\cite{ORTW92}. The quantum theory as it currently stands is detailed
in \cite{BS92}. Such algebras for the  affine Toda theories are
discussed in e.g. \cite{FHM92} and other references in \cite{OSU93}.

The above methods have the similarity that they quantise the whole
of the theory. Perhaps of more relevance for the soliton solutions
is the approach of Hollowood \cite{Ho93}, where the quantum theory
of the $\hat A_r$ abelian Toda solitons is assumed to be encapsulated in their
S-matrix, which can be conjectured and shown to satisfy a
restrictive set of constraints. This is much the same thing as is
done for the Toda particles. A further exciting possibility for
soliton quantisation appears in the paper \cite{PPZ93}. Here, the
authors consider an N=2 supersymmetric field theory. They construct
a combination of the spinor charges whose square is zero and can
thus be used as a BRST-type operator for the quantum theory. It
turns out that the energy-momentum tensor is a variation under the
symmetry generated by this charge, and so correlation functions
become independent of position in space-time. Furthermore the
particle excitations all become unphysical leaving only topological
degrees of freedom. This seems very similar to our setting the
free-field to zero and would thus be interesting to investigate
further in the present context.

As well as the quantum theory, there is also the interesting
possibility of further investigation at the classical level. We
would like to rule out all multisoliton solutions with real energies
other than those given. It is probable that this happens if we
restrict all of the rest of the conserved charges to be real, given
some normalisation. These charges have been given for the $\hat A_r$ abelian
theories in \cite{Ni92}.

It is hoped that the non-abelian Toda theories will receive the
attention that they appear to deserve, and that has already been
focussed on the abelian theories.

\section{Acknowledgements}

I am grateful to David Olive and Peter Bowcock for helpful
discussion, and to Patrick Dorey for discussions and pointing out
the correct version of a conjecture in the original version of this
paper. I am also indepted particularly to Mikhail Saveliev for
taking the time to explain the theories he is partly responsible for
originating. Finally I especially thank the UK Science and Engineering
Research Council for financial support.
\section{Structure Constants of the Kac-Peterson Basis}\label{kpb}
\label{constants}
In this appendix we indicate how the coefficients
$q_\alpha^\pm$ appearing in the commutation relation \ref{defq} and
the energy-momentum formula \ref{multisol} can be calculated.

First let us project down from the loop algebra into the Lie algebra
$\fing$, by setting the formal parameter $t$ to unity, for example. The
image of $\hat X_\pm$ under this projection will be elements
$h_\pm\in H_R$, with eigenvalues $e^{\pm 2\pi i/m}$
under conjugation by $S_w$. The precise definition of these elements
was discussed in section 3. The structure constants $q^\pm_\alpha$
are given by \beq q^\pm_\alpha=\alpha(h_\pm).\label{qexp}\eeq All we
need to know, besides the root system of $\fing$, is an expansion of
$h_\pm$ over the generators of $H_0$. In fact since we know the way
that $H_R$ is embedded into $H_0$ we only need to know the expansion
of $h_\pm$ over the generators of $H_R$. This problem has been
solved in \cite{FLO91}, where the authors use the properties of the
Coxeter element discovered by Kostant \cite{Ko59} to expand the
images of $h_\pm$ under the canonical isomorphism with $H_R^\ast$
over the simple roots of the algebra. Crucial to this procedure is
the bicolouration property of finite type Dynkin diagrams; their
points can be labelled `black' or `white' with no points of like colour
adjacent. Defining $\delta_jB=1$ if $i$ is black and zero otherwise
yield the following formula, valid for the simple roots of a simple
algebra: \beq q^+_{\alpha_j}=2ie^{-i\delta_{jB}\theta}\sin\theta
x_j(1),\label{qformula}\eeq  where $\theta={\pi\over h}$, and $x_j(1)$
are the components of the right Perron-Frobenius vector of the Cartan
matrix of $\fing$, this being the eigenvector of least eigenvalue.
The formula in this form was lifted from \cite{OTU93}.

The expression \ref{qformula} can easily be generalised to the case
where $\fing_R$ is not semisimple, the constants $q^\pm$ then being
an appropriately normalised sum of such terms.

\subsection{Vanishing Structure Constants}

When $w$ is not a Coxeter element of the Weyl group of $\fing$ there
will generally be some $w$-orbits of roots $\alpha$ which are
orthogonal to $h_\pm$. We can see from the form of the solution
given in \ref{fulsolred}, for example, that the $\exp(Q\faz)$ for
such roots lead to trivial solutions of the equations of motion. It
seems reasonable to suppose that these group elements should be
omitted altogether as not contributing anything to the soliton
spectrum, merely shifting the field $h$ by a constant group element,
whose Adjoint action on $\hat X_\pm$ is trivial. We get a trivial
solution whatever the value of $Q$. This leads us to the conclusion
that the existence of such $\faz$ implies that the constant, vacuum,
solutions of the theory are not a discrete set. This should not come
as any surprise if we recall the remarks of section 4, about some
irritating portions of the conserved currents. The projection
operator $P$ when applied to the equation of motion \ref{eqmn} gives
\beq \partial_-P(h^{-1}\partial_+h)=0.\label{rubbish}\eeq The
existence of this rather trivial portion of the field was first
noted by Leznov and Saveliev in the original description
\cite{LS83}. It seems that this cannot simply be factored out of the
theory since the motion of the rest of the fields will still depend
on the solution of \ref{rubbish}. It could well be that there will
be a valid interpretation of these fields, but this awaits further study.

\section{Vertex Operators in other Representations}
\label{otherrep}
In this appendix we indicate, along much the same lines as for the
principal vertex operator construction in \cite{OU93}, how the
existence of vertex operator representations for the $\faz$ in the
level one representations of simply-laced affine algebras allows us
to deduce properties of the $\faz$ in the other representations.
These are necessary for the construction of the matrix elements in
the Leznov-Saveliev solution \ref{lssol}. First let us describe how
the results can be extended to the rest of the fundamental
representations of the simply-laced algebras (and we know that the
general highest weight representations can be constructed by taking
symmetrised tensor product of these, though these representations
are not really necessary for our purposes).

We expect that all of the fundamental representations can be found
in the decompositions of tensor products of the level one
representations (known henceforth as the minimal representations).
In principle then, we could use the vertex operator construction in
such a tensor product, and then decompose to obtain the fundamental
representation. In practice this computation would seem to be
difficult, but it can be done (see the simple example
of $\hat D_4$ in \cite{OU93} for example). However, we do not really
need to perform these calculations as the explicit form of the
solution is of no real interest to us (see the remarks in the
discussion). What is interesting from our point of view is the
asymptotic behaviour of the solution, which is governed by the
greatest non-vanishing power of $\faz$ within a representation. We
have already seen that the square of $\faz$ must vanish within the
minimal representations. How can we use this to answer the above
question?

Notice that a representation of level $x$ can only be found
in the tensor product of $x$ minimal representations.\footnote{Note
that the construction by tensor products of a given representation
is not necessarily unique.} This is
obvious from the fact that $x$ is the eigenvalue of $K$. Within such
a representation the maximum non-vanishing power of $\faz$ must be
$\leq x$, since any greater power would have all of its constituent terms
with a quadratic or greater power acting on some fundamental
representation. I am unable to prove the equality, although it seems
unlikely that any power of $\faz$ not actually required to vanish
should do so. We shall thus assume that the greatest non-vanishing
power of $\faz$ in a level $x$ representation of a simply-laced
$\gh$ is $x$.

In this paper we are particularly interested in the representation
with highest weight $\rhc$, for the purpose of evaluating the
energy and momentum of the soliton solutions. The level of this
representation is $m$, since \beq x=\rhc(K)=(d',K)=m,\eeq from the
definition of $d'$.

Now we discuss the value of vertex operator constructions for the
non-simply laced algebras. In \cite{OU93} it was noted that every
non-simply laced algebra, including the twisted algebras, can be obtained as a
subalgebra of a
simply-laced algebra invariant under the canonical lift of one of its diagram
automorphisms. Such automorphisms are of course outer. The value of
this observation was that the highest weight representation theory
of the simply-laced algebra, and in
particular the principal vertex operator method for constructing an arbitrary
such representation, can be used to describe the representation
theory of the $\faz$. In particular it turns out that the $\faz$ for
the non-simply laced algebra are just invariant linear combinations
of those for the simply-laced algebra. In  the present, more general,
case, given a Weyl group element $w$ (and equivalently a regular
semisimple subalgebra $\fing_R$) of a non-simply laced algebra
$\fing$ embedded in a
simply-laced algebra $\fing^\supset$, we see that there will be a
corresponding $\fing^\supset_R$ as the embedded image of $\fing_R$,
which is preserved by the diagram automorphism.  A little thought
along the same lines as in \cite{OU93} reveals that we expect the
$\faz$ of the non-simply laced  loop algebra to again be linear
combinations of the $\faz$ of the simply-laced loop algebra, since
we can in fact be precise about the action of the automorphism on
the $\hat E_I$. This means that the simply-laced $\faz$ provide a permutation
representation of the diagram automorphisms (up to signs); to get the
non-simply
laced $\faz$ we just identify the trivial one-dimensional
representations. For more detail on all of these points again refer to
\cite{OU93}.

Finally if the non-simply laced $\faz$ is a linear combination of
$y$ simply-laced $\faz$, then its $xy+1^{\rm th}$ power must vanish
in a representation of level $x$.

There are other approaches to the construction of representations of
the non-simply-laced affine algebras. The extension of the
Kac-Peterson construction to the twisted algebras was discussed in
the conclusion of the original article \cite{KP85}. This would presumably   be
useful to present and discuss twisted non-abelian affine Toda
theories. An alternative approach to the construction of homogeneous
vertex operator constructions was found in \cite{GNOS86} where the
authors found that the sum of $\faz$ can actually be expressed as a
single vertex operator if we introduce some fermionic Fock space as
well as that of the homogeneous Heisenberg subalgebra. It is
possible that this sort of approach could be used for the more
general $\faz$ discussed here.
\section{On the Centre of $Z(\hxp)$}\label{centre}

In the discussion of the abelianisation procedure in section 4, for
example in equation \ref{cA}, the
local conserved charges which can be obtained by this method were
shown to correspond to the elements of the centre of $Z(\hxp)$.
There is a natural conjecture for the form of these elements, given
the procedure of section \ref{ccem} for constructing an appropriate
$\hxp$. We proceed in exactly the same way as in the appendix of
\cite{OT2}. First consider the element $h_+\in\fing$ of which $\hxp$
is the grade 1 affinisation. If an element of the loop algebra
$\tilde \fing$ belongs to the
centre of $Z(\hxp)$ then it follows that its natural projection into
$\fing$, for example by setting the formal parameter $t$ to unity,
belongs to the centre of $Z(h_+)$.

Because $h_+$ is an
eigenvector of the conjugation by $S_w$, with eigenvalue $e^{2\pi
i/m}$, it follows that $Z(h_+)$ is a $\Bbb C$-span of eigenvectors of
the conjugation. This in turn means that its centre is also of this
form. Let $C\in \fing\otimes\fing$ by the Casimir tensor of $\fing$,
invariant under the adjoint action of $\fing$ on
$\fing\otimes\fing$. It is easy to show that elements of $\fing$ of the
form \beq h^{(n)}=\Tr_L h^nC\label{cencen}\eeq must lie in the
centre of $H(h_+)$. $\Tr_L$ denote the trace in some representation
of the left hand factor of the tensor product. Note that $h^{(n)}$ has
eigenvalue $e^{2\pi
in/m}$ under the conjugation. It seems reasonable to suppose that
the non-zero $h^{(n)}$ actually span the whole of the centre. This
conjecture could perhaps be verified using results in \cite{Ko59}
and \cite{Dy57}.

\section{The Dual Weyl Representation} \label{rhoproof} In this appendix we
discuss
the properties of a particular highest weight representation of the
algebra $\gh$, which we shall call the dual Weyl representation.

Focus on the element $x$ which defines the grading of $\gh$. We note
that $x\in \fing_R$ is semisimple (ad-diagonalisable), which means
that there must be at least one Cartan subalgebra $H_\bigtriangleup$
of $\fing$ which contains it, and this Cartan subalgebra is clearly
contained in $\fing_0$. We can make a particular choice for
$H_\bigtriangleup$ by using the fact that there is a {\em unique}
Cartan subalgebra of $\fing_R$ which commutes with $x$ (because $x$
is a regular element of $\fing_R$). Call this $H_R'$ and fix \beq
H_R'\oplus H_\bot=H_\bigtriangleup\subset \fing_0\subset\gh_0. \eeq
We can guarantee (see e.g. \cite{Hu72}) that $x$ will lie in the
dominant Weyl chamber of some system of simple roots   (strictly the
image of this chamber under the canonical isomorphism between
$H_\bigtriangleup^\ast$ and $H_\bigtriangleup$). The advantage of
this choice is that the step-operators $e_\alpha$ corresponding to
the positive roots will have non-negative grades, and those
corresponding to the negative roots will have non-positive grades.
Implementing the Kac-Peterson procedure we get a correspondingly
$d'$-graded basis $t^ie_\alpha$ of the loop algebra. This shows that
we can legitimately define highest weight representations with
respect to the $d'$-gradation, i.e. representations generated from a
vector $|\Lambda\rangle $ which is annihilated by all of the step
operators of positive grade, and which satisfies \beq
h_\bigtriangleup|\Lambda\rangle= \Lambda(h_\bigtriangleup)|
\Lambda\rangle \forall h_\bigtriangleup\in
H_\bigtriangleup\subset\gh \label{defhwt}\eeq

This picture of the Kac-Peterson basis as being the affinisation of
a step-operator basis defined with respect to a different
Cartan subalgebra of $\fing$ allows us to be more specific about the
action of $\gh_0$ on the highest weight representations defined in
this way. Suppose that $h_a\in H_\bigtriangleup$ is the Cartan element
corresponding to each simple root $a$ of the Kac-Moody algebra
$\gh$. It is well known \cite{Kac90} that there exists a highest
weight representation for which the $\Lambda(h_a)$ are equal to any
set of non-negative integers. In particular there must exist a representation
with
highest weight state $|\rhc\rangle$ with $(d',h_a)=\rhc(h_a)$ since
\beq (d',h_a)=(d',[e_a,e_{-a}])= ([d',e_a],e_{-a})\eeq are all
non-negative integers.

We need to know the action of $\gh_0$ on the highest weight state.
Orthogonally decompose $\gh_0=H_\bigtriangleup\oplus \gh_\bigtriangleup$, where
$\gh_\bigtriangleup$ is a direct sum of step operators with $d'$-grades zero:
if it includes a step operator for a particular root $b$ it also
includes that for $-b$. Because $(d',h_b)=\rhc(h_b)=0$ we know that
neither of $\rhc\pm b$ can be weights of the representation. Thus we
conclude that $|\rhc\rangle$ is annihilated by $\gh_\bigtriangleup$. This
representation is put to use in section \ref{emsol}.


\begin{thebibliography}{10}

\bibitem{OSU93}
D.I. Olive, M.V. Saveliev, and J.W.R. Underwood.
\newblock On the solitonic specialisation of some two-dimensional completely
  integrable systems.
\newblock {\em preprint Imperial/TP/92-93/13 NI92014 SWAT/4}, 1992.

\bibitem{LS92}
A.N. Leznov and M.V. Saveliev.
\newblock {\em Group-Theoretical Methods for Integration of Nonlinear Dynamical
  Systems}, volume~15 of {\em Progress in Physics}.
\newblock Birkhauser-Verlag, Basel, 1992.

\bibitem{OTU93}
D.I. Olive, N.~Turok, and J.W.R. Underwood.
\newblock Solitons and the energy-momentum tensor for affine {T}oda theory.
\newblock {\em Nucl. Phys. B, to appear}, 1993.

\bibitem{FLO91}
A.~Fring, H.C. Liao, and D.I. Olive.
\newblock The mass spectrum and coupling in affine {T}oda field theory.
\newblock {\em Phys. Lett.}, B266:82, 1991.

\bibitem{Do91}
P.E. Dorey.
\newblock Root systems and purely elastic {S}-matrices.
\newblock {\em Nucl. Phys.}, B358:654, 1991.

\bibitem{Do92}
P.E. Dorey.
\newblock Root systems and purely elastic {S}-matrices 2.
\newblock {\em Nucl. Phys.}, B374:741, 1992.

\bibitem{DGZ92}
G.W. Delius, M.T. Grisaru, and D.~Zanon.
\newblock Exact {S}-matrices for non-simply-laced affine {T}oda theories.
\newblock {\em Nucl. Phys.}, B382:365, 1992.

\bibitem{BCDS90}
H.W. Braden, E.~Corrigan, P.~E. Dorey, and R.~Sasaki.
\newblock Affine {T}oda field theory and exact {S-M}atrices.
\newblock {\em Nucl. Phys.}, B338:465, 1990.

\bibitem{FO92}
A.~Fring and D.I. Olive.
\newblock The fusing rule and scattering matrix of affine {T}oda theory.
\newblock {\em Nucl. Phys.}, 379B:429, 1992.

\bibitem{LS83}
A.N. Leznov and M.V. Saveliev.
\newblock Two-dimensional exactly and completely integrable dynamical systems
  etc.
\newblock {\em Comm. in Math. Phys.}, 89:59, 1983.

\bibitem{Ho92}
T.J. Hollowood.
\newblock Solitons in affine {T}oda field theories.
\newblock {\em Nucl. Phys.}, page 523, 1992.

\bibitem{MM92}
N.J. MacKay and W.A. McGhee.
\newblock Affine {T}oda solitons and automorphisms of {D}ynkin diagrams.
\newblock {\em preprint DTP-92-45/RIMS-890}, 1992.

\bibitem{ACFGZ92}
H.~Aratyn et. al.
\newblock Hirota's solitons in the affine and conformal affine {T}oda model.
\newblock {\em preprint UICHEP-TH-92-18}, 1992.

\bibitem{BB92}
O.~Babelon and D.~Bernard.
\newblock Affine solitons.
\newblock {\em preprint SPhT-92-055, LPTHE-92-16.}, 1992.

\bibitem{OU93}
D.I. Olive, N.~Turok, and J.W.R. Underwood.
\newblock Affine toda solitons and vertex operators.
\newblock {\em preprint Imperial/TP/92-93/29, SWAT/92-93/5, PUP-TH-93/1392},
  1993.

\bibitem{Dy57}
E.~B. Dynkin.
\newblock Semisimple subalgebras of semisimple {L}ie algebras.
\newblock {\em AMS Translations, Series 2}, 6:111, 1957.

\bibitem{Ko59}
B.~Kostant.
\newblock The principal three-dimensional subgroup and the {B}etti numbers of a
  complex simple {L}ie group.
\newblock {\em American Journal of Mathematics}, 81:973, 1959.

\bibitem{BS92}
P.~Bouwknegt and K.~Schoutens.
\newblock {W-S}ymmetry in conformal field theory.
\newblock {\em preprint CERN-TH.6583/92, ITP-SB-92-23}, 1992.

\bibitem{ORTW92}
L.~O'Raifeartaigh, P.~Ruelle, I.~Tsutsui, and A.~Wipf.
\newblock {W-A}lgebras for generalised {T}oda theories.
\newblock {\em Comm. in Math. Phys.}, 143:333, 1992.

\bibitem{FWBFO89}
P.~Forg\'acs et. al.
\newblock Liouville and {T}oda theories as conformally reduced {WZNW} theories.
\newblock {\em Phys. Lett.}, 227:214, 1989.

\bibitem{KP85}
V.G. Kac and D.H. Peterson.
\newblock 112 constructions of the basic representation of the loop group of
  {E$_8$}.
\newblock In {\em Proceedings of the Conference `Anomalies, Geometry,
  Topology', Argonne}. World Scientific, March 1985.

\bibitem{Hu72}
J.E. Humphreys.
\newblock {\em Introduction to {L}ie Algebras and Representation Theory}.
\newblock Springer, 1972.

\bibitem{Ca72}
R.W. Carter.
\newblock Conjugacy classes in the Weyl group.
\newblock {\em Compositio Mathematica}, 25:1, 1972.

\bibitem{GO86}
P.~Goddard and D.I. Olive.
\newblock {K}ac-{M}oody and {V}irasoro algebras in relation to quantum physics.
\newblock {\em Int. J. of Mod. Phys}, A1:303, 1986.

\bibitem{OT2}
D.I. Olive and N.~Turok.
\newblock Local conserved densities and zero-curvature conditions for {T}oda
  lattice field theories.
\newblock {\em Nucl. Phys.}, 257B [FS15]:277, 1985.

\bibitem{GORS93}
J.-L.~Gervais et. al.
\newblock Gauge conditions for the constrained-{WZNW}-{T}oda reductions.
\newblock {\em Phys. Lett.}, B301:41, 1993.

\bibitem{Bo89}
P.~Bowcock.
\newblock Canonical quantisation of the gauged {W}ess-{Z}umino model.
\newblock {\em Nucl. Phys.}, B316:80, 1989.

\bibitem{GS92}
J.-L. Gervais and M.V. Saveliev.
\newblock Black holes from non-abelian {T}oda theories.
\newblock {\em Phys. Lett.}, B286:271, 1992.

\bibitem{BB91}
O.~Babelon and D.~Bernard.
\newblock Dressing transformations and the origin of the quantum group
  symmetries.
\newblock {\em Phys. Lett.}, B220:81, 1991.

\bibitem{Ba88}
O.~Babelon.
\newblock Extended conformal algebra and the {Y}ang-{B}axter equation.
\newblock {\em Phys. Lett.}, B215:523, 1988.

\bibitem{ABBP93}
E.~Aldrovandi, L.~Bonora, V.~Bonservizi, and R.~Paunov.
\newblock Free field representation of {T}oda field theories.
\newblock {\em Preprint SISSA 210/92/EP}, 1993.

\bibitem{FHM92}
L.~Feh\'er, J.~Harnad, and I.~Marshall.
\newblock Generalised {D}rinfeld-{S}okolov reductions and {KdV} type
  hierarchies.
\newblock {\em preprint UdeM-LPN-TH-92/103, CRM-1832}, 1992.

\bibitem{Ho93}
T.J. Hollowood.
\newblock Quantising $sl(n)$ solitons and the {H}ecke algebra.
\newblock {\em preprint OUTP-29-03P}, 1992.

\bibitem{PPZ93}
S.~Penati, M.~Pernici, and D.~Zanon.
\newblock Solitons in two-dimensional topological field theories.
\newblock {\em preprint IFUM-441-FT}, 1993.

\bibitem{Ni92}
M.R. Niedermeyer.
\newblock The spectrum of the conserved charges in affine {T}oda theory.
\newblock {\em preprint DESY 92-105}, 1992.

\bibitem{GNOS86}
P.~Goddard, W.~Nahm, D.I. Olive, and A.~Schwimmer.
\newblock Vertex operators for non-simply laced algebras.
\newblock {\em Comm. in Math. Phys.}, 107:179, 1986.

\bibitem{Kac90}
V.G. Kac.
\newblock {\em Infinite-dimensional {L}ie algebras}.
\newblock Cambridge University Press, 1990.

\end{thebibliography}
\end{document}